\documentclass[aps,twocolumn,showpacs,preprintnumbers,nofootinbib,prl,superscriptaddress,10pt]{revtex4-1}

\usepackage[utf8]{inputenc}
\usepackage{graphicx}
\usepackage{amsmath,amssymb}
\usepackage{amsfonts}
\usepackage{physics}
\usepackage{xspace}
\usepackage{bm}
\usepackage[normalem]{ulem}
\usepackage{mathrsfs}
\usepackage[dvipsnames]{xcolor}
\usepackage[unicode]{hyperref}
\hypersetup{colorlinks=true, citecolor=MidnightBlue,
	linkcolor=CornflowerBlue, urlcolor=Blue, linktocpage=true}
\usepackage[all]{hypcap}
\usepackage{multirow}
\usepackage[format=plain,justification=RaggedRight,labelsep=endash,font=small,labelfont=sc]{caption}
\usepackage[export]{adjustbox}

\definecolor {darkgreen}{rgb}{0.2,0.7,0.2}

\definecolor {dark}{rgb}{0.43,0.5,0.5}

\newcommand{\eq}{\begin{equation}}
\newcommand{\be}{\begin{equation}}
\newcommand{\eeq}{\end{equation}}
\newcommand{\ee}{\end{equation}}

\begin{document}

\title{Dynamics of Screening in Modified Gravity}
 \author{Lotte ter Haar}
 \affiliation{SISSA, Via Bonomea 265, 34136 Trieste, Italy and INFN Sezione di Trieste}
 \affiliation{IFPU - Institute for Fundamental Physics of the Universe, Via Beirut 2, 34014 Trieste, Italy}
 \author{Miguel Bezares}
 \affiliation{SISSA, Via Bonomea 265, 34136 Trieste, Italy and INFN Sezione di Trieste}
 \affiliation{IFPU - Institute for Fundamental Physics of the Universe, Via Beirut 2, 34014 Trieste, Italy}
 \author{Marco Crisostomi}
 \affiliation{SISSA, Via Bonomea 265, 34136 Trieste, Italy and INFN Sezione di Trieste}
 \affiliation{IFPU - Institute for Fundamental Physics of the Universe, Via Beirut 2, 34014 Trieste, Italy}
  \author{Enrico Barausse}
 \affiliation{SISSA, Via Bonomea 265, 34136 Trieste, Italy and INFN Sezione di Trieste}
 \affiliation{IFPU - Institute for Fundamental Physics of the Universe, Via Beirut 2, 34014 Trieste, Italy}
 \author{Carlos Palenzuela}
\affiliation{Departament  de  F\'{\i}sica,  Universitat  de  les  Illes  Balears  and  Institut  d'Estudis Espacials  de  Catalunya,  Palma  de  Mallorca,  Baleares  E-07122,  Spain}
\affiliation{Institut Aplicacions Computationals (IAC3),  Universitat  de  les  Illes  Balears,  Palma  de  Mallorca,  Baleares  E-07122,  Spain}

\begin{abstract}
Gravitational theories differing from General Relativity  may explain the  accelerated expansion of the Universe without  a cosmological constant. However, to pass local gravitational tests, a ``screening mechanism''
is needed to suppress, on small scales, the fifth force driving the cosmological acceleration.
We consider the simplest of these theories, i.e. a scalar-tensor theory with first-order derivative self-interactions, and study isolated (static and spherically symmetric) non-relativistic and relativistic stars. We produce screened solutions
and use them as initial data for non-linear numerical evolutions in spherical symmetry.
We find that these solutions are stable under large initial perturbations, as long as they do not cause gravitational collapse. When gravitational collapse is triggered,
the characteristic speeds of the scalar evolution equation diverge, even before
apparent black-hole or sound horizons form.
This  casts doubts on whether the dynamical evolution of screened stars may be predicted in these effective field theories.
\end{abstract}

\date{\today \hspace{0.2truecm}}

\maketitle
\flushbottom

\textit{Introduction.}---The accelerated expansion of the universe is among the biggest mysteries of cosmology. While
achievable  by a cosmological constant or a Dark-Energy (DE) component, these possibilities face long-standing theoretical issues~\cite{Weinberg:1988cp}.
The possibility that 
cosmic acceleration may arise from a modification of General Relativity (GR) on cosmological scales has thus attracted considerable attention \cite{Clifton:2011jh}.

The simplest extension of GR is provided by scalar-tensor theories, where gravity is described not only
by two tensor polarizations, but also by a scalar graviton. Their most general form is given by degenerate higher-order scalar-tensor theories~\cite{BenAchour:2016fzp},
which contain well-known examples such as Fierz-Jordan-Brans-Dicke (FJBD) theory~\cite{Fierz:1956zz,Jordan:1959eg,Brans:1961sx}, dilatonic Gauss-Bonnet theory~\cite{Kobayashi:2011nu},  Horndeski~\cite{Horndeski:1974wa}
and beyond-Horndeski~\cite{Gleyzes:2014dya} theories, etc.

While scalar-tensor theories can produce self-accelerated cosmic expansion without a 
cosmological constant \cite{Crisostomi:2017pjs}, they typically  produce also local deviations from GR on small scales~\cite{Berti:2015itd}.
These include the solar system and binary pulsars, where GR has been tested to exquisite accuracy~\cite{Damour:1991rd,Will:1993hxu,Will:2014kxa},
and the compact-object binaries observed by gravitational-wave (GW) interferometers~\cite{LIGOScientific:2019fpa}. 
However, some theories possess ``screening mechanisms'' (Vainshtein screening~\cite{Vainshtein:1972sx,Babichev:2013usa}, $k$-mouflage~\cite{Babichev:2009ee}, chameleon/symmetron
screening~\cite{Khoury:2003rn,Hinterbichler:2010es}, etc.) that locally
produce  a GR-like phenomenology, potentially passing existing constraints.
Screening has only been tested in static/quasi-static configurations, but its validity is often taken for granted also in dynamical settings, e.g. GW generation \cite{Belgacem:2019pkk}.
Here, we will verify this assumption. 

We consider scalar-tensor theories with first-order derivative self-interactions ($k$-essence \cite{Chiba:1999ka, ArmendarizPicon:2000dh}). Among the many theories aiming to explain DE, $k$-essence is among the few  unconstrained by the GW170817 bound on the GW speed~\cite{Monitor:2017mdv,TheLIGOScientific:2017qsa}, and by other constraints based on GW propagation~\cite{Creminelli:2018xsv, Creminelli:2019kjy, Babichev:2020tct}. 
By studying static and spherically symmetric solutions, we confirm the presence of ``kinetic'' screening ($k$-mouflage \cite{Babichev:2009ee}) in non-relativistic stars, and 
extend it to fully relativistic, compact stars. 
We then consider spherically  symmetric time evolutions of these screened solutions,
 using the fully non-linear code of~\cite{us_long_paper}.
 
\textit{Static spherically symmetric screening.}---With  units $\hbar=c=1$
and  signature $(-+++)$, the $k$-essence action in the Einstein frame is
\be
S=\int \mathrm{d}^{4}x\sqrt{-\tilde{g}}\left[\frac{M_{\mathrm{Pl}}^{2}}{2}\tilde{R} + K(\tilde{X})  \right]  + S_{m}\left[g_{\mu\nu},\Psi\right] \,. \label{action}
\ee
Here, $M_{\mathrm{Pl}}=(8\pi G)^{-1/2}$ is the Planck mass;
$g_{\mu\nu}=A(\phi) \tilde{g}_{\mu\nu}$ and 
$\tilde{g}_{\mu\nu}$ are respectively 
the metrics in the Jordan and Einstein frames; the conformal factor is $A(\phi)=e^{\alpha \, \phi /M_{\mathrm{Pl}}}$, where $\alpha\sim {\cal O}(1)$ is dimensionless;
$\tilde{g}$ and $\tilde{R}$ are the (Einstein-frame) metric determinant and  Ricci scalar; $\tilde{X}\equiv \tilde{\nabla}_\mu \phi \tilde{\nabla}^\mu \phi$ is the standard kinetic term of the scalar field $\phi$.
Variation of the
action yields
\begin{align}\label{fieldeqs1}
&\tilde{G}_{\mu\nu}=8 \pi G\big[K(\tilde{X})\tilde{g}_{\mu\nu}-2K'(\tilde{X})\tilde{\nabla}_\mu \phi \tilde{\nabla}_\nu \phi + \tilde{T}_{\mu\nu}\big]\,,\\\label{fieldeqs2}
&\tilde{\nabla}_\mu\left(K'(\tilde{X}) \tilde{\nabla}^\mu \phi\right)=\frac{1}{4}A^{-1}(\phi)A'(\phi)\tilde{T}\,,
\end{align}
where $\tilde{G}_{\mu\nu}$ and $\tilde{T}_{\mu\nu}$ are respectively the  Einstein and energy-momentum tensors in the Einstein frame\footnote{While the Einstein frame is convenient to solve 
the equations numerically (see e.g. \cite{Barausse:2012da,Palenzuela:2013hsa}), when presenting results (unless stated otherwise) we transform back to the Jordan frame, where the physical interpretation is clearer; e.g. we plot results vs the Jordan-frame areal radius $r$, which differs by a conformal factor from the Einstein-frame radius $\tilde{r}$.}.

For $K(\tilde{X})$ we consider only the lowest-order terms
\be
K(\tilde{X}) = - \frac12 \tilde{X} + \frac{\beta}{4\Lambda^4}\tilde{X}^{2} - \frac{\gamma}{8\Lambda^8}\tilde{X}^{3} \,, \label{Kform}
\ee
with $\Lambda$ the strong-coupling scale of the effective field theory (EFT), and $\beta, \gamma\sim {\cal O}(1)$ dimensionless coefficients.
We assume that the background scalar field is responsible for DE, therefore $\Lambda \sim (H_0 M_{\mathrm{Pl}})^{1/2}\sim 5\times 10^{-3}$ eV, where $H_0$ is the present-day Hubble expansion rate.
It is exactly the hierarchy $M_{\mathrm{Pl}} \gg \Lambda$, needed for cosmology, that allows for screening local scales (a much larger $\Lambda$ would make 
$k$-essence equivalent to FJBD, where no screening appears).
Screened solutions are possible for any $\beta<0,\,\gamma>0$ \footnote{Without gravity, this choice leads to superluminal propagation and non-analytic $2 \to 2$ scattering amplitudes in the forward limit \cite{Adams:2006sv}, but see~\cite{Alberte:2020jsk} for recent developments.},
but in the following we set $\beta = 0$ and $\gamma=1$, which
ensures that $1 +  2\, X\, K''(X)/K'(X) >0$ for all $X$ (a sufficient condition to avoid Tricomi-type breakdowns of the Cauchy problem~\cite{us_long_paper}; see also \cite{Babichev:2007dw, Brax:2014gra}). Our conclusions also hold for more general $\beta$ and $\gamma$, if they are such that this condition holds (see examples  in~\cite{us_long_paper}).

In more detail, \cite{Babichev:2009ee} suggested that non-relativistic  stars in $k$-essence present a $k$-mouflage  mechanism, whereby GR is recovered within a ``screening radius'' $r_{\rm k}\sim\Lambda^{-1}\sqrt{M/M_{\rm Pl}}$ (with $M$ the star's mass), as a result of the non-linear terms in Eq.~\eqref{Kform} dominating over the linear one\footnote{This non-linear regime may seem problematic from an EFT view-point. However, \cite{deRham:2014wfa} (without gravity) and \cite{Brax:2016jjt} (with gravity) showed that quantum corrections are under control in the non-linear regime.}. To check this, we first consider constant-density, non-relativistic stars.
Using the same weak-field approximation  applied in~\cite{Babichev:2010jd,Babichev:2013pfa} to study screening in massive (bi-)gravity,
we obtain an approximate
equation for the scalar-field radial derivatives $y\equiv\phi'$ and $y'$ (with $'\equiv {\rm d}/{\rm d}{r}$): 
{\small 
\begin{align}\label{y_eq}
& \frac{r \rho }{M_{\mathrm{Pl}}\Lambda ^2 } = y' \Bigg[\frac{\left(3 \alpha
   ^2+2\right) r}{\alpha  \Lambda ^2}+\frac{r^2 y}{M_{\mathrm{Pl}}\Lambda ^2 } +\frac{15\gamma r y^4 (2M_{\mathrm{Pl}}+\alpha r y)}{4M_{\mathrm{Pl}}\alpha \Lambda^{10}}\Bigg] \nonumber \\
& + \frac{2 \left(3
   \alpha ^2+2\right) y}{\alpha  \Lambda ^2}+\frac{2 \alpha ^2 r y^2}{M_{\mathrm{Pl}}\Lambda ^2 } +\frac{3 \gamma y^5}{\alpha  \Lambda
   ^{10}} +\frac{3 \gamma  r y^6}{M_{\mathrm{Pl}}\Lambda
   ^{10}} +\frac{5 \alpha  \gamma  r^2 y^7}{4M_{\mathrm{Pl}}^2 \Lambda ^{10}
   } \,.
\end{align}
}

Approximate analytic solutions to this equation can be obtained in the stellar interior: $y_1\approx [\alpha \rho r \Lambda^8/(3\gamma M_{\mathrm{Pl}})]^{1/5}$; in the exterior within the screening radius: $y_2\approx[\alpha M \Lambda^8/(4\pi\gamma M_{\mathrm{Pl}}r^2)]^{1/5}$; and outside the screening radius: $y_3\approx \mathrm{const}/r^2$.
In the FJBD case $\beta=\gamma=0$, an approximate solution is given by $y_3$ outside the star, and by $y_0\approx \alpha \rho r/[2 M_{\mathrm{Pl}}(2+3\alpha^2)]$ inside.

These approximate solutions show that in $k$-essence
the scalar derivative (which encodes the additional ``fifth force'' beyond GR) 
is suppressed inside $r_{\rm k}$. However, the inner solution is not regular at the star's center. Regularity requires $y=\phi'\propto r$ when $r\to 0$, and a different behavior is not acceptable, as it would cause the appearance of a central conical singularity. 

To amend this behavior, we solve numerically Eq.~\eqref{y_eq}, imposing $y\to 0$ when $r\to 0$  as a boundary condition. This completely determines
the solution as Eq.~\eqref{y_eq} does not involve $y''$. Thus, it is not trivial that the regular solution will match the approximates ones ($y_1,\,y_2,\,y_3$) above. In more detail, since Eq.~\eqref{y_eq} is singular at $r=0$, we must solve it perturbatively at small radii, 
imposing   $y \propto r$ when $r\to 0$. This yields another approximate solution,
(which at leading order matches the approximate FJBD inner solution
$y_0$) which we use
to ``inch away'' from $r=0$ and provide initial conditions for the numerical integration. 
\begin{figure*}
\begin{center}
\includegraphics[width=\textwidth]{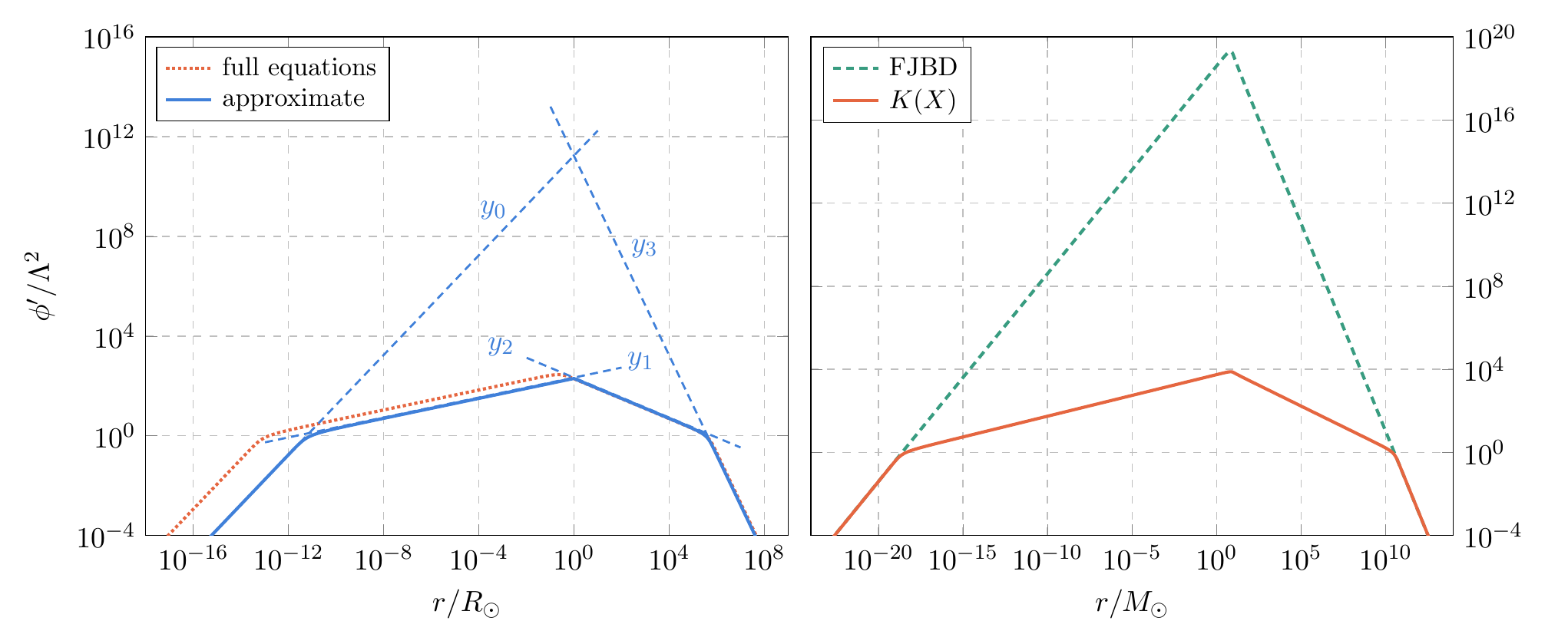}
\caption{$\phi'$  vs $r$ for $\gamma=1$ and $\alpha=1/2$. \textit{Left:} A weakly gravitating, Sun-like star.
We plot the numerical solution of Eq.~\eqref{y_eq} (solid blue line), the approximate solutions $y_0$, $y_1$, $y_2$, $y_3$ (dashed blue lines), and the numerical solution of the full system  \eqref{fieldeqs1}-\eqref{fieldeqs2} (dotted orange line).
\textit{Right:} A neutron star in $k$-essence (solid orange line) and FJBD  ($\beta=\gamma=0$, dashed green line).}
\label{scalar}
\end{center}
\end{figure*}
This procedure gives the numerical solution (regular at the center) shown by a solid line in Fig.~\ref{scalar} (left panel), where we also compare to the approximate solutions $y_0$, $y_1$, $y_2$ and $y_3$. As can be seen, the regular numerical solution
matches the approximates solutions $y_1$, $y_2$ and $y_3$ everywhere but near the center, where we find agreement with $y_0$ (the FJBD solution) instead.

These results confirm the existence of (regular) $k$-mouflage solutions in non-relativistic stars, but it is not obvious that the same will apply to strongly gravitating relativistic stars, e.g. neutron stars, or even for weakly gravitating stars when 
the full system \eqref{fieldeqs1}--\eqref{fieldeqs2} is solved simultaneously.
We therefore
write Eqs.~\eqref{fieldeqs1}--\eqref{fieldeqs2}
using a spherically symmetric ansatz for the (Einstein-frame) metric $\mathrm{d}\tilde{s}^2=\tilde{g}_{tt}(\tilde{r}){\rm d} t^2+\tilde{g}_{\tilde{r}\tilde{r}}(\tilde{r}){\rm d} \tilde{r}^2+ \tilde{r}^2{\rm d} \Omega^2$
and for the scalar field, and solve the coupled system by imposing regularity at the center. Since  Eqs.~\eqref{fieldeqs1}--\eqref{fieldeqs2} depend on $\phi$ (and not only on $\phi'$ and $\phi''$, unlike Eq.~\eqref{y_eq}), an additional boundary condition is needed for $\phi$. We thus require  $\phi$ to approach a constant $\phi_\infty$ at spatial infinity. 
If we take $|\phi_\infty|/\Lambda\lesssim 1$, as expected from cosmological considerations, results
are robust against the exact value of $\phi_\infty$.

We adopt a polytropic equation of state $p=K \rho_b^\Gamma$, $p=(\Gamma-1) (\rho-\rho_b)$ --  with $p,\,\rho\,,\rho_b,$ the pressure, energy density and baryonic density -- in the Jordan frame (thus, the equation of state in the Einstein frame involves the conformal factor, c.f. \cite{Barausse:2012da,Palenzuela:2013hsa}).
We use  $K=123\;G^3 M_\odot^2/c^6$ and $\Gamma=2$ for neutron stars, and $K=5.9\times10^{-3}\;G^{1/3}R^{2/3}_\odot/c^{2/3}$ and $\Gamma=4/3$ for weakly gravitating, Sun-like stars. 
We impose regularity by solving perturbatively the equations near the center, and use this solution to provide initial conditions for the outbound integration at small but non-zero $r$. These initial conditions depend on the central values of the scalar field and density. We fix the former via a shooting procedure by requiring  $\phi\to\phi_{\infty}$
as $r\to\infty$, while the central density is varied on a grid to produce stars of different masses.

The solution
 for a Sun-like star is shown
 in Fig.~\ref{scalar} (left panel, dotted orange line), and presents the same 
qualitative features as the approximate solution obtained previously.
Similarly, the radial profile of $\phi'$ for neutron stars (right panel, solid orange line)
 shows kinks right outside the center, at the stellar surface, and at the screening radius. We also plot by a dashed green line the solution to Eq.~\eqref{fieldeqs1}-\eqref{fieldeqs2} obtained
for $\beta=\gamma=0$ (i.e. FJBD). The 
$k$-mouflage solution matches the FJBD one near the center and outside $r_{\rm k}$, but deviates from it (suppressing $\phi'$ and thus the scalar force) when non-linearities become important (i.e. when $X/\Lambda^4\gtrsim 1$). 
Similar plots and conclusions apply to generic 
$\beta<0$ and $\gamma>0$.

\begin{figure*}
\begin{center}
\includegraphics[width=\textwidth]{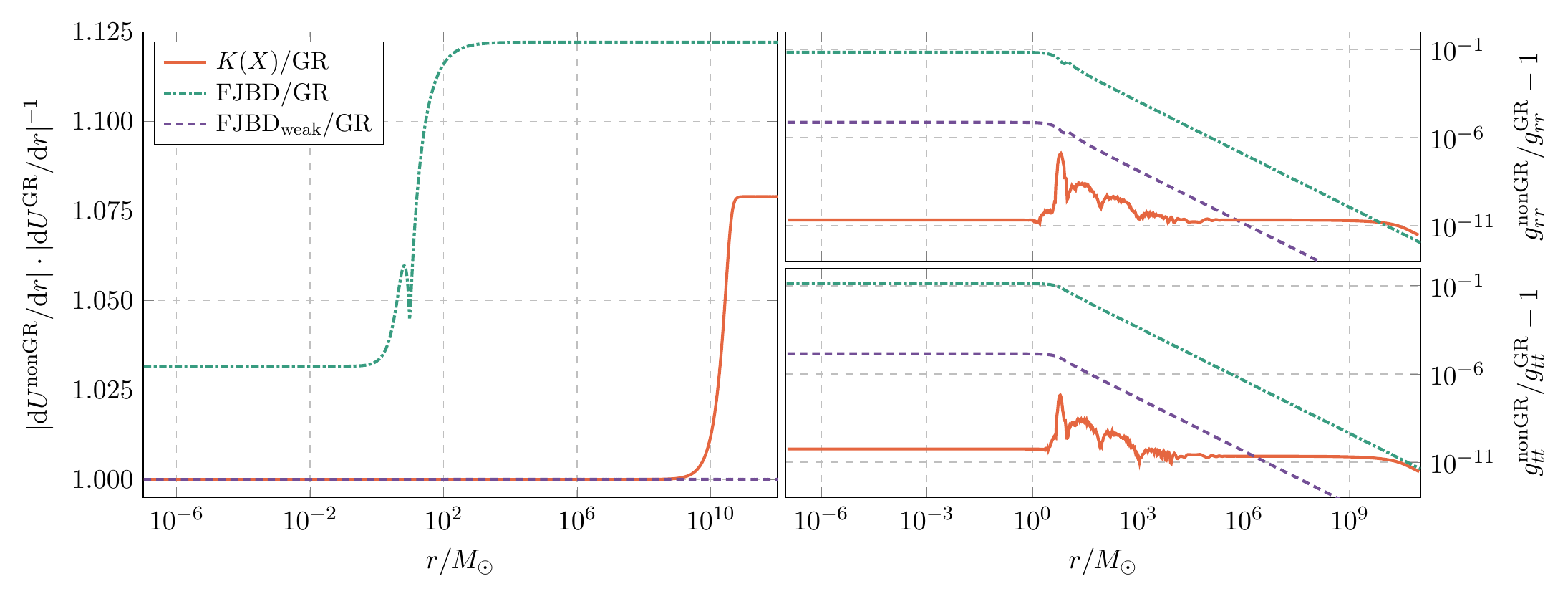}
\caption{Deviations of the metric and its derivatives from GR, for
$k$-essence with $\gamma=1$ and $\alpha=1/2$, and for
FJBD (with $\alpha=1/2$ and $\alpha=5\cross10^{-3}$, the latter referred to as  $\mathrm{FJBD}_{\mathrm{weak}}$).
 \textit{Left:} Ratio of the Jordan-frame Newtonian forces 
 with the GR counterpart. The FJBD star is slightly smaller than in GR, which explains the feature at $r/M_\odot\sim10$.
 \textit{Right:} The fractional deviation from GR of the $g_{rr}$ (upper panel) and $g_{tt}$ (lower panel) components of the Jordan-frame metric.}
\label{metric}
\end{center}
\end{figure*}

Again for neutron stars, in Fig.~\ref{metric} (left panel) we show the ratio of the Newtonian force $|{\rm d} U/{\rm d} r|$, with $U= -(g_{tt}+1)/2$ the Newtonian potential, 
for solutions in $k$-essence and FJBD theory with respect to solutions in GR, as a function of the Jordan-frame areal radius $r$. Note that the scalar-field contribution (fifth force) is suppressed in $k$-essence relative to FJBD theory inside $r_{\rm k}$, as expected from screening. 
In Fig.~\ref{metric} (right panel) we also show the 
fractional deviations of the (Jordan-frame) metric components $g_{tt}$ and $g_{rr}$ 
from GR, in FJBD theory (with $\alpha=1/2$ and $\alpha=5\times 10^{-3}$) and in $k$-essence  (with $\alpha=1/2$). Note that the tiny deviations from GR in $k$-essence suggest that not only is the Newtonian dynamics  essentially equivalent to GR's, but that the same holds also at  first post-Newtonian order. 
This is apparent from the comparison with FJBD theory with 
$\alpha=5\times 10^{-3}$, which is in agreement 
with current solar system tests of the post-Newtonian dynamics~\cite{Will:1993hxu,Will:2014kxa,Barausse:2012da,Palenzuela:2013hsa}.

\begin{figure}
        \centering
        \includegraphics[width=0.45\textwidth]{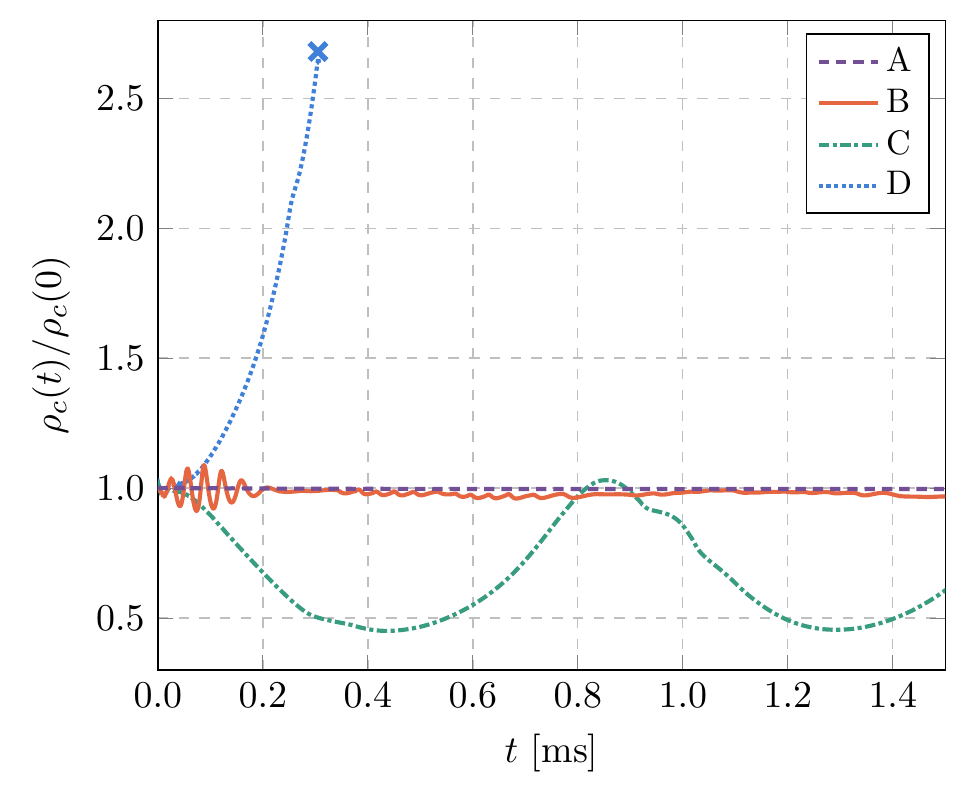}
        \caption{Evolution of the (Einstein-frame) central density of a $k$-mouflage star for $\Lambda\simeq10^{6}\;\mathrm{eV}$ (with $\alpha\simeq0.2$ and $\gamma=1$), for unperturbed initial data (A); small initial perturbations (B); large  perturbations that initially decrease (C)/increase (D) the star's compactness. Case D leads to  collapse and diverging characteristic speeds (at the time marked by a cross).
        }
        \label{fig_ev}
\end{figure}

\textit{Screening perturbations and time evolutions.}---To check the stability of our static spherical solutions,
we  numerically evolve the scalar, the metric and the matter fields according to
Eqs.~\eqref{fieldeqs1}--\eqref{fieldeqs2}. 
We employ the 1+1 (i.e. spherically symmetric
but time-dependent) fully non-linear evolution code used in~\cite{us_long_paper}
for the vacuum case, supplementing it with matter as described in \cite{ValdezAlvarado:2012xc}.
Both the matter's and the scalar's evolution are expressed as conservation laws and  integrated with high-resolution shock-capturing (HRSC) methods.
We first checked that if  static spherical solutions
(for both Sun-like and neutron stars) are used as initial data, the system does not evolve (e.g. case A in Fig.~\ref{fig_ev})\footnote{This  is not trivial. Even for these initial data, numerical evolutions 
 break with standard finite-difference or even soft shock-capturing methods,
 presumably  as a result of strong micro-shocks in the scalar field~\cite{Reall:2014sla,Babichev:2016hys,Bernard:2019fjb}, which form even from smooth initial data.
 This suggests that Eqs.~\eqref{fieldeqs1}--\eqref{fieldeqs2} only allow for weak solutions (i.e. solutions to the integral version of the system), which we successfully
 obtain  by using HRSC methods.}. However, if we perturb them (in their matter or scalar content),
the results vary dramatically according to $\Lambda$ and the perturbation amplitude/sign.

For $\Lambda \gtrsim 10^{7}\;\mathrm{eV}$, the static spherical initial data show no screening
and are very similar to FJBD theory, as expected. Non-linearities in the scalar sector are never excited and evolutions are well-behaved, however large the initial perturbations. 
For screened solutions ($\Lambda \lesssim 10^{6}\;\mathrm{eV}$), the outcome of time evolutions depends  on the initial perturbation amplitude/sign.
Small perturbations (case B in Fig.~\ref{fig_ev}) and large ones initially decreasing the stellar compactness (case C in Fig.~\ref{fig_ev}) oscillate but do not grow, confirming the stability of the screened solutions. 
However, when large perturbations have the right sign to  trigger gravitational collapse (case D in Fig.~\ref{fig_ev}), the characteristic propagation speeds of the scalar-field equation eventually diverge, even before apparent/black-hole horizons form.
In more detail, the characteristic speeds are encoded in the principal part (i.e.
the part involving only the highest derivatives) of 
Eq.~(\ref{fieldeqs2}), which is given by $\gamma^{\mu\nu}\tilde\nabla_\mu \tilde\nabla_\nu \phi$,
with $\gamma^{\mu\nu} \equiv \tilde{g}^{\mu\nu} + 2[{\,K''(\tilde X)}/{K'(\tilde X)}]\tilde{\nabla}^{\mu}\phi\tilde{\nabla}^{\nu}\phi$. Writing
the principal part in first-order form, i.e. $\partial_t \boldsymbol{U}+
\boldsymbol{V}\partial_r \boldsymbol{U}$, with $\boldsymbol{U}\equiv(\partial_t\phi,\partial_r\phi)$ and $\boldsymbol{V}$ the characteristic matrix, the characteristic speeds are then the eigenvalues of $V$~\cite{us_long_paper}:
\begin{eqnarray}
v_\pm = -\frac{ \gamma^{tr}}{ \gamma^{tt}}\pm \sqrt{\frac{-{\rm det}( \gamma^{\mu\nu})}{( \gamma^{tt})^2}} \,.
\end{eqnarray}
Their non-linear divergence, appearing because $\gamma^{tt} \to 0$,  is known to plague $k$-essence also in vacuum (for initial data close to critical collapse)~\cite{us_long_paper,Bernard:2019fjb,Figueras:2020dzx}, and resembles that of the Keldysh equation $t\,\partial^2_t\phi(t,r) + \partial^2_r\phi(t,r) = 0$, which is hyperbolic with characteristic speeds $\pm(-t)^{-1/2}$ for $t<0$.
The problem persists when looking for screened solutions through relaxation of GR stars, as done in \cite{Barausse:2012da}.

As we stressed in~\cite{us_long_paper}, diverging characteristic speeds are not necessarily pathological and may occur  because of gauge choices (see e.g. a wave equation on flat space in Eddington-Finkelstein coordinates, ${\rm d}s^2=-{\rm d} v^2+2 {\rm d}v {\rm d}r+r^2 {\rm d}\Omega^2$). Like in vacuum~\cite{us_long_paper},
the characteristic speeds may be kept finite during the evolution if  a non-vanishing shift in the metric is allowed. Nevertheless, because of the non-linear nature of the field equations, we could not identify 
a suitable coordinate condition (i.e. a choice of lapse and/or shift) avoiding these divergences and simultaneously producing stable evolutions at least in 1+1 dimensions.
We tried different shift conditions that successfully keep the velocities finite, but those still lead to unstable evolutions even with HRSC methods.
Whatever its interpretation (physical or due to the gauge), 
the divergence of the characteristic speeds is troublesome in practice.
The Courant-Friedrichs-Lewy stability condition implies that the timestep $\Delta t$
of a numerical evolution should be $\Delta t<\Delta r/v$, with $\Delta r$ the spatial resolution  and $v$ the maximum characteristic speed. Clearly, $\Delta t\to0$ as $v\to\infty$, i.e. simulations must
grind to a halt when the characteristic speeds diverge. 

Therefore, our results suggest that the field equations of $k$-essence cannot be 
evolved starting from $k$-mouflage solutions.
A possible practical solution to  evolve the dynamics of $k$-mouflage may consist of using implicit methods~\cite{Pareschi:2010,Palenzuela:2008sf}. However, the latter may not recover the system's true dynamics, as they might miss the small-timescale features of the solution (and their cumulative secular effect, if any). In other words, implicit methods integrate
out the ultraviolet (UV) details of the solution, which might be crucial to achieve a well-posed Cauchy evolution.

\textit{Conclusions.}---We have shown that kinetic screening ($k$-mouflage) of scalar effects occurs in isolated stars in $k$-essence, even when the stars are highly compact/relativistic and the physically important
requirement of regularity at the star's center is accounted for.
$k$-mouflage solutions are stable to small perturbations, and also to large ones as long as they do not cause gravitational collapse. However, when  large perturbations with the right sign to trigger collapse are applied to  $k$-mouflage solutions, the evolution leads to diverging characteristic speeds for the scalar, well before the formation of apparent black-hole/sound horizons. This divergence might not be pathological in itself, but prevents dynamical evolutions of the  collapse of $k$-mouflage stars. $k$-essence thus loses predictability on $k$-mouflage configurations subject to these large perturbations. 
This is a serious  flaw, as the theory cannot make predictions about the general time-dependent evolution of stars (including their collapse to a black hole), at least in 1+1 dimensions. This is markedly different than in GR~\cite{Shibata_2000,Pretorius:2005gq} or FJBD~\cite{Barausse:2012da}, where spherical dynamical simulations of compact objects present no such problems. If kinetic screening exists, $k$-essence is therefore (at best) incomplete in general dynamical settings.
A UV-completion of $k$-essence may render the time evolution of screened stars well-posed.
However,
it is not guaranteed that $k$-mouflage solutions will still be present in such UV completions, see e.g. \cite{Burrage:2020bxp}. Moreover, positivity bounds suggest that
locality and/or Lorentz-invariance may have to be violated to
UV-complete the theory~\cite{Adams:2006sv}, if screening solutions are to be present.
One may however attempt to modify the theory's equations in a UV-agnostic way inspired by dissipative hydrodynamics, possibly allowing for successfully evolving the dynamics~\cite{Allwright:2018rut}.
\begin{acknowledgments}
{\em Acknowledgments.}---L.t.H, M.B, M.C, and E.B. acknowledge support from the EU's H2020 ERC Consolidator Grant
``GRavity from Astrophysical to Microscopic Scales'' (Grant No.  GRAMS-815673). 
C.P. acknowledges support from the Spanish Ministry of Economy and Competitiveness Grants No. AYA2016-80289-P and No. PID2019-110301GB-I00 (AEI/FEDER, UE). We thank J. M. Ibáñez, L. Lehner and G. I. Montecinos for enlightening conversations on the well-posedness of the Cauchy problem.

\end{acknowledgments}

\bibliographystyle{apsrev4-1}
\bibliography{master}

\end{document}